\documentclass[aps,prapplied,amsmath,amssymb,twocolumn,superscriptaddress,nofootinbib]{revtex4-2}

\usepackage[english]{babel}
\usepackage{graphicx}
\usepackage{svg}
\usepackage{adjustbox}
\usepackage{fixltx2e}
\usepackage{makecell}
\usepackage{hyperref}
\usepackage{cleveref}
\usepackage{subcaption}
\usepackage{caption} 
\graphicspath{ {./pictures/} }
\usepackage{overpic}
\usepackage{siunitx}

\begin{document}


\title{CMOS-Compatible, Wafer-Scale Processed Superconducting Qubits Exceeding Energy Relaxation Times of 200µs}


\author{T. Mayer*}
\email{thomas.mayer@emft.fraunhofer.de}
\affiliation{Fraunhofer Institut für Elektronische Mikrosysteme und Festkörpertechnologien EMFT, Munich, Germany}

\author{J. Weber*}
\email{johannes.weber@emft.fraunhofer.de}
\affiliation{Fraunhofer Institut für Elektronische Mikrosysteme und Festkörpertechnologien EMFT, Munich, Germany}

\author{E. Music}
\affiliation{Fraunhofer Institut für Elektronische Mikrosysteme und Festkörpertechnologien EMFT, Munich, Germany}

\author{C. Morán Guizán}
\affiliation{Fraunhofer Institut für Elektronische Mikrosysteme und Festkörpertechnologien EMFT, Munich, Germany}

\author{S. J. K. Lang}
\affiliation{Fraunhofer Institut für Elektronische Mikrosysteme und Festkörpertechnologien EMFT, Munich, Germany}

\author{L. Schwarzenbach}
\affiliation{Fraunhofer Institut für Elektronische Mikrosysteme und Festkörpertechnologien EMFT, Munich, Germany}

\author{C. Dhieb}
\affiliation{Fraunhofer Institut für Elektronische Mikrosysteme und Festkörpertechnologien EMFT, Munich, Germany}

\author{B. Kiliçlar}
\affiliation{Fraunhofer Institut für Elektronische Mikrosysteme und Festkörpertechnologien EMFT, Munich, Germany}

\author{A. Maiwald}
\affiliation{Fraunhofer Institut für Elektronische Mikrosysteme und Festkörpertechnologien EMFT, Munich, Germany}

\author{Z. Luo}
\affiliation{Technical University of Munich, Munich, Germany}

\author{W. Lerch}
\affiliation{Fraunhofer Institut für Elektronische Mikrosysteme und Festkörpertechnologien EMFT, Munich, Germany}

\author{D. Zahn}
\affiliation{Fraunhofer Institut für Elektronische Mikrosysteme und Festkörpertechnologien EMFT, Munich, Germany}

\author{I. Eisele}
\affiliation{Fraunhofer Institut für Elektronische Mikrosysteme und Festkörpertechnologien EMFT, Munich, Germany}
\affiliation{Center Integrated Sensor Systems (SENS), Universität der Bundeswehr München, Munich, Germany}

\author{R.N. Pereira}
\affiliation{Fraunhofer Institut für Elektronische Mikrosysteme und Festkörpertechnologien EMFT, Munich, Germany}

\author{C. Kutter}
\affiliation{Fraunhofer Institut für Elektronische Mikrosysteme und Festkörpertechnologien EMFT, Munich, Germany}
\affiliation{Center Integrated Sensor Systems (SENS), Universität der Bundeswehr München, Munich, Germany}

\collaboration{*These authors contributed equally to this work and are listed in alphabetical order.}




\date{\today}

\begin{abstract}
We present the results of an industry-grade fabrication of superconducting qubits on 200\,mm wafers utilizing CMOS-established processing methods. By automated waferprober resistance measurements at room temperature, we demonstrate a Josephson junction fabrication yield of $\SI{99.7}{\percent}$ (shorts and opens) across more than 10000 junctions and a qubit frequency prediction accuracy of $\SI{1.6}{\percent}$. In cryogenic characterization, we provide statistical results regarding energy relaxation times of the qubits with a median $T_1$ of up to 100\,µs and individual devices consistently approaching 200\,µs in long-term measurements. This represents the best performance reported so far for superconducting qubits fabricated by industry-grade, wafer-level subtractive processes.
\end{abstract}



\maketitle

\section{Introduction}
The industry-grade fabrication of superconducting qubits has gained increasing interest recently. The usage of large wafers ($\geq$200\,mm) and the leveraging of methods established in the semiconductor industry could prove to be one of the main drivers for the necessary scaling of superconducting quantum processing units (QPUs). Especially, the utilization of alternative methods for Josephson junction fabrication has been discussed to replace the
widespread angled evaporation and lift-off approach, where on-wafer process gradients and the presence of organic photoresist during junction formation can complicate parameter targeting and introduce loss sources, respectively\cite{mohseni2025buildquantumsupercomputerscaling,ke2025scaffoldassistedwindowjunctionssuperconducting,Megrant2025-si}. The successful wafer-scale fabrication of superconducting qubits within industry-grade pilot lines has already been demonstrated, utilizing a subtractive processing approach, where the bottom and top electrodes of the JJs are patterned by subsequent deposition, optical lithography, and etching of two Aluminum layers under vacuum breaking\cite{lang2025advancingsuperconductingqubitscmoscompatible,Van_Damme2024-vv}. In this article, we present the results from a fabrication run applying the process presented in Ref. \cite{lang2025advancingsuperconductingqubitscmoscompatible}, but advanced by additional optimization measures. Across eight 200\,mm wafers used in the run with $>$10000 investigated qubits, we achieved a room-temperature JJ yield of $\approx\SI{99.7}{\percent}$ (shorts and opens). Within the best split group, we observed a median $T_1$ of 100\,µs, with a minimum value of 20\,µs and the maximum exceeding 200\,µs for both $T_1$ and $T_2^{\mathrm{echo}}$ times in cryogenic analysis of multiple chips, representing the highest $T_1$ times reported so far for qubits fabricated by industry-grade, wafer-level subtractive processing.

\section{Methods and chip design}\label{sec:methods}
For the results presented in the article, the same chip design, main process sequence, and characterization approach was applied as presented in Ref. \cite{lang2025advancingsuperconductingqubitscmoscompatible}, where more details can be found. The following provides a brief summary.\\
\textbf{Wafer fabrication.} The eight wafers processed in the presented run were fabricated in the 200\,mm pilot line at Fraunhofer EMFT. A high-ohmic Si-substrate (3-5\,k$\Omega$cm) is HF-dipped and coated with the first Al layer via a sputtering process. For patterning the base layer, resonators, and the bottom electrode of the JJ, an optical i-line stepper lithography and plasma-assisted dry etching is applied. Next, in an in-situ sequence, the native oxide on the bottom Al layer is removed by Ar sputtering, the JJ tunneling oxide is formed, and the second Al layer sputtered to encapsulate the formed oxide. The vacuum is then broken and the second Al layer is again patterned by optical i-line lithography and dry-etching to form the top electrode of the JJ.\\
\textbf{Wafer-level electrical characterization.} After successful fabrication, the wafers are electrically characterized by fully automated room-temperature waferprober measurements. Additional to the probing of a variety of test structures placed in the wafers' dicing streets, the resistance of the Qubits' Josephson junction on the chips is directly measured by placing a probing needle on each pad of the shunt capacitor and generating an I-V-curve.\\
\textbf{Dicing and Bonding.} To prepare single chips for the cryogenic characterization, the wafers are covered by a protection resist and diced using a diamond saw. Under consideration of the results from the wafer-level electrical characterization, specific chips are picked from the wafers, the protection resist removed by acetone, isopropyl alcohol, and DI water rinsing, and wire-bonded with Aluminum wires to a cryo-compatible printed circuit board (PCB).\\
\textbf{Chip-level cryogenic characterization.} The chips were mounted in a Bluefors LD400 dilution refrigerator capable of achieving base temperatures around 10\,mK. The refrigerator is equipped with over 60 RF and 40 DC lines to facilitate high-throughput measurements. To enhance experimental efficiency and flexibility, cryogenic microwave switches were integrated at the mixing chamber (MXC) stage, allowing dynamic signal routing to increase experimental efficiency.
Initial qubit characterization was conducted in the frequency domain using one- and two-tone spectroscopy. These measurements enabled identification of the readout resonator frequencies as well as the fundamental qubit's transition frequency and anharmonicity. Subsequent time-domain experiments were performed using Zurich Instruments' Quantum Computing Control System (QCCS). Each qubit underwent a comprehensive calibration sequence comprising Rabi amplitude scans to determine optimal drive strengths and Ramsey interferometry to fine-tune qubit frequencies. Further assessments included measurements of energy relaxation times ($T_1$), free induction decay times ($T_2^*$), and Hahn echo coherence times ($T_2^{\mathrm{echo}}$), providing insights into decoherence mechanisms. For selected devices, long-term stability studies were conducted to evaluate temporal fluctuations in coherence properties over extended periods, which are critical for reliable quantum information processing.\\
\textbf{Chip design.} For benchmarking of processing quality, a chip design is used hosting four fixed-frequency, floating pocket Transmon qubits, where a single Josephson junction is placed between two isolated shunt capacitor pads surrounded by the superconducting base layer and capacitively coupled to a resonator that is used to control and read-out the qubit. The qubits are designed with a target frequency of $\SI{4.42}{\giga\hertz}$. The resonator itself is capacitively coupled to a feedline that features pads for wire-bonding on each side to connect the chip to the PCB. The chip has a size of 4.3 $\times$ 7\,mm$^2$ and each wafer fits up to 750 chips, which equals 3000 qubits, plus additional test structures. However, in the fabrication run presented in this article, for every second chip on each wafer an alternative test design was used that is not included in the discussed results.\\

\section{Results}
\subsection{Electrical room temperature characterization and frequency targeting}

\begin{figure*}[htbp]
  \centering
  \begin{minipage}[t]{0.48\linewidth}
    \begin{overpic}[width=\linewidth]{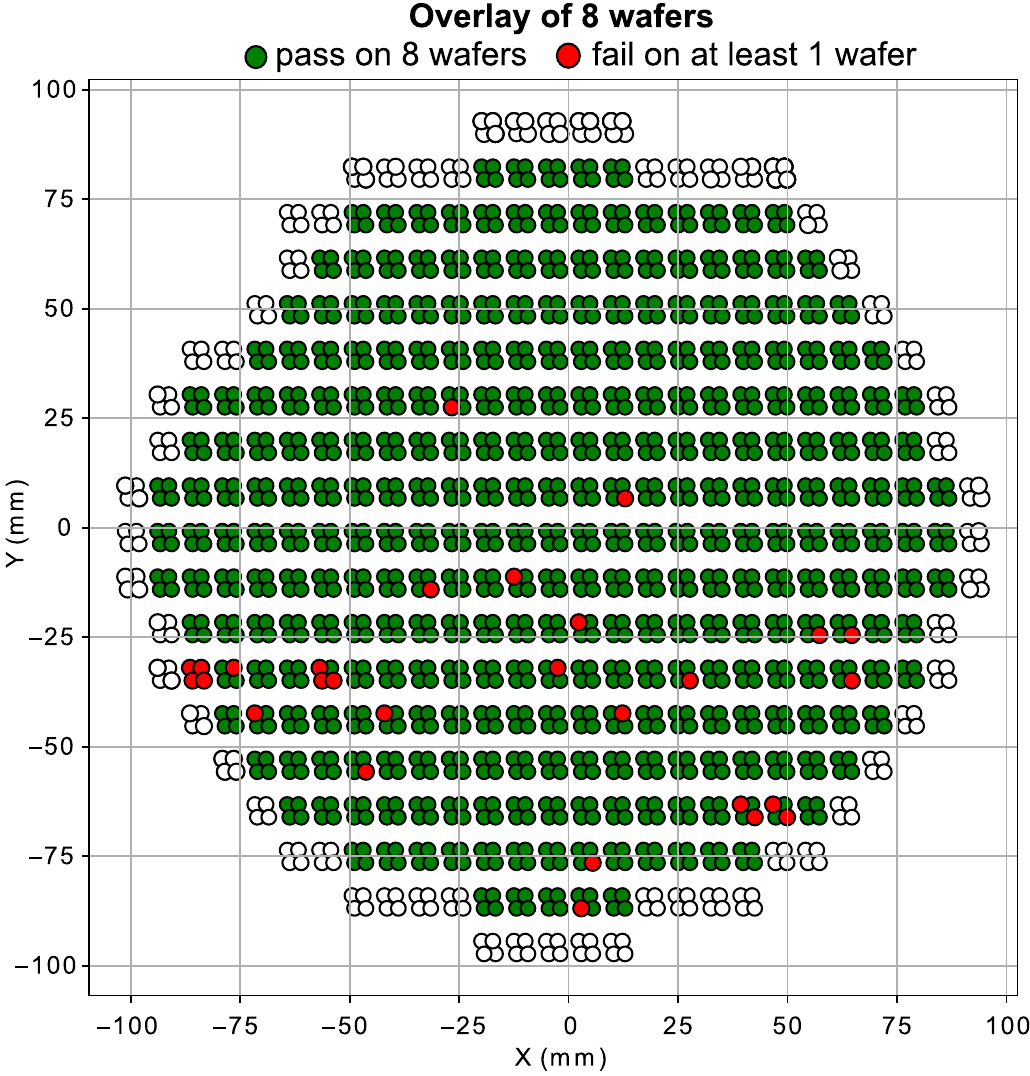}
      \put(13,87){\textbf{a)}} 
    \end{overpic}
  \end{minipage}
  \hfill
  \begin{minipage}[t]{0.48\linewidth}
    \begin{overpic}[width=\linewidth]{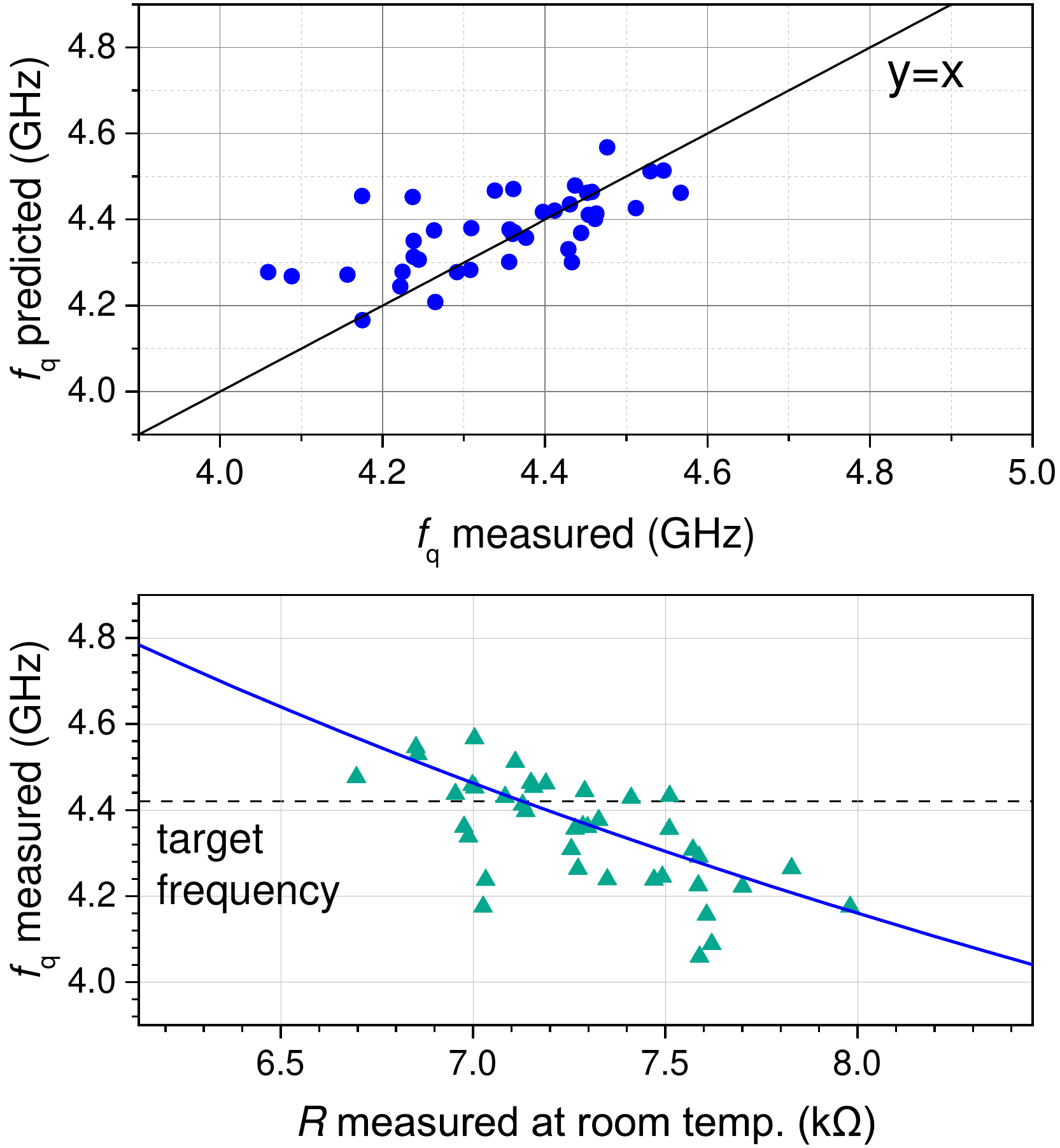}
      \put(14,96){\textbf{b)}}
      \put(14,45){\textbf{c)}}
    \end{overpic}
  \end{minipage}
  \caption{a) Overlay wafermap of 8 wafers showing a qubit's Josephson junction as fail (red), when one junction across the eight wafers does not fulfill the pass criteria and green, when it passes on all eight. b) Qubit frequency predicted from normal state junction resistance via Eq. \ref{eq:baratoff} versus the cryogenically measured frequency. c) Cryogenically measured qubit frequency vs. normal state junction resistance $R_n$ with the design frequency of $\SI{4.42}{\giga\hertz}$ highlighted as a black dashed line and Ambegaokar-Baratoff prediction as a blue line.}
  \label{fig:platzhalter1}
\end{figure*}

Electrical wafer-level characterization at room temperature provides an indispensable tool for process control monitoring and quality assessment of the fabrication process with high statistical significance. Especially the direct probing of the qubits' JJs, as explained in Sec. \ref{sec:methods}, proofed as a powerful predictor of qubit yield in our previous studies \cite{Lang2023-hi, lang2025advancingsuperconductingqubitscmoscompatible, lang2025waferscalecharacterizationalalxoyaljosephson, mayer20253dintegratedsuperconductingqubitscmoscompatible}. Figure \ref{fig:platzhalter1}a) shows a pass/fail map of the Qubits' JJs on each chip overlaying the results of all eight wafer investigated in the discussed processing run. Within the map, a JJ is counted as \textit{fail}, when the room temperature resistance measurement yielded a short ($R < \SI{1}{\kilo\ohm}$) or an open ($R > \SI{50}{\kilo\ohm}$). Each circle represents the JJ of one qubit and hence each group of four circles represents one four-qubit chip. A circle is colored red (fail), when the specific JJ failed on at least one of the eight wafers and only appears green (pass), when it was \textit{pass} on all 8 wafers. An edge exclusion is applied to disregard the outermost chips at the wafer edge (white circles), since they are typically less reliable as the processing is affected by the chips not having an outside neighbor. Across the eight wafers, the resistances of 10072\footnote{The number regards the applied edge exclusion and the fact that every second chip on the wafers used an alternative test design, which is not part of the article.} qubit JJs were measured and a total of 30 fails were detected. This room temperature JJ yield of $\SI{99.7}{\percent}$ shows the high precision, repeatability, and reliability of the applied subtractive processing approach and underlines the promise of the method for future scaling of Transmon-type QPUs to larger qubit numbers, as well for the application in alternative architectures, like e.g. Fluxoniums, where the series connection of a large number of JJs demands particular processing reliability.\\
The room temperature resistance measurement of the qubits' JJs not only provides a reliable indicator for low temperature qubit yield, but can also be used as a predictor for qubit frequency $f_{01}$ via the Ambegaokar-Baratoff model \cite{ambegaokar_tunneling_1963}
\begin{equation}\label{eq:baratoff}
    f_{\text{01}} = \sqrt{\frac{1}{R_{\mathrm{n}}}\frac{0.882k_{\mathrm{B}}T_{\mathrm{c}}}{hC_{\mathrm{q}}}}-\frac{e^2}{2hC_{\mathrm{q}}},
\end{equation}
where $f_{01}$ denotes the transition frequency of the ground to the first excited state of the qubit, $R_n$ the normal state resistance of the JJ measured at room temperature, $C_{\mathrm{q}}$ the qubit capacity, $T_{\mathrm{c}}$ the critical temperature of the superconductor, and $k_{\mathrm{B}}$, $h$, and $e$ the Boltzman and Planck constants as well as the electron charge, respectively. Assuming the design qubit capacity of $C_{\mathrm{q}}=\SI{86}{\femto\farad}$ and $T_{\mathrm{c}}=\SI{0.71}{\kelvin}$ as obtained from previous studies\cite{lang2025advancingsuperconductingqubitscmoscompatible}, Fig. \ref{fig:platzhalter1}b) shows the predicted frequency from Eq. \ref{eq:baratoff} versus the cryogenically determined actual qubit frequency of the selected chips picked from multiple wafers of the processing run. The black line with slope 1 acts as a guide to the eye representing perfect prediction accuracy. Across 40 evaluated qubits, the  frequency predicted at room temperature deviates by $\SI{1.6}{\percent}$ from the cryogenically measured frequency on average\footnote{The value comprises both uncertainties of the room- and low-temperature measurements, as well as deviations of each qubit from the design capacity value $C_{\mathrm{q}}=\SI{86}{\femto\farad}$, which was assumed for the calculation. Hence, a further improvement is possible by measurement accuracy optimizations and especially by including a determined JJ capacitance value for each separate qubit.}. 
Utilizing such precise frequency forecasting, we can use Eq. \ref{eq:baratoff} to define a target normal state resistance corresponding to the design frequency of the qubit. For our chip design, where a frequency of $\SI{4.42}{\giga\hertz}$ is implemented, a target resistance of $R_{\mathrm{n}}^{\mathrm{target}}\approx \SI{7.1}{\kilo\ohm}$ is obtained. From the waferprober measurements, chips can now be identified, where the JJs of all four qubits lie in the vicinity of $R_{\mathrm{n}}^{\mathrm{target}}$. Applying the arbitrary specification limits of $\pm$10\% around $R_{\mathrm{n}}^{\mathrm{target}}$, the resulting frequency targeting is shown in Fig. \ref{fig:platzhalter1}c). Without applying any post-processing of the JJs, like e.g. laser\cite{Hertzberg2021-yn, doi:10.1126/sciadv.abi6690, 10.1063/5.0102092} or THz annealing to adjust $R_n$, the average deviation from the design frequency lies at $\approx$2.5\%\footnote{It should be noted that all four qubits on each chip are always considered in the statistics, not just the best value on each chip.}. Obviously, this targeting could be further improved by applying tighter specification limits below $\pm$10\% around $R_{\mathrm{n}}^{\mathrm{target}}$. While this reduces the number of chips passing the limits, the high amount of chips on large-scale wafers still can provide sufficient yield for very tight limits. Since frequency targeting might pose a crucial bottleneck to scale to large QPU systems\cite{PhysRevResearch.5.043001}, the wafer-scale fabrication approach can provide an additional upside in this regard.

\subsection{Qubit performance}

\begin{figure*}[htbp]
    \centering
    \begin{overpic}[width=\textwidth]{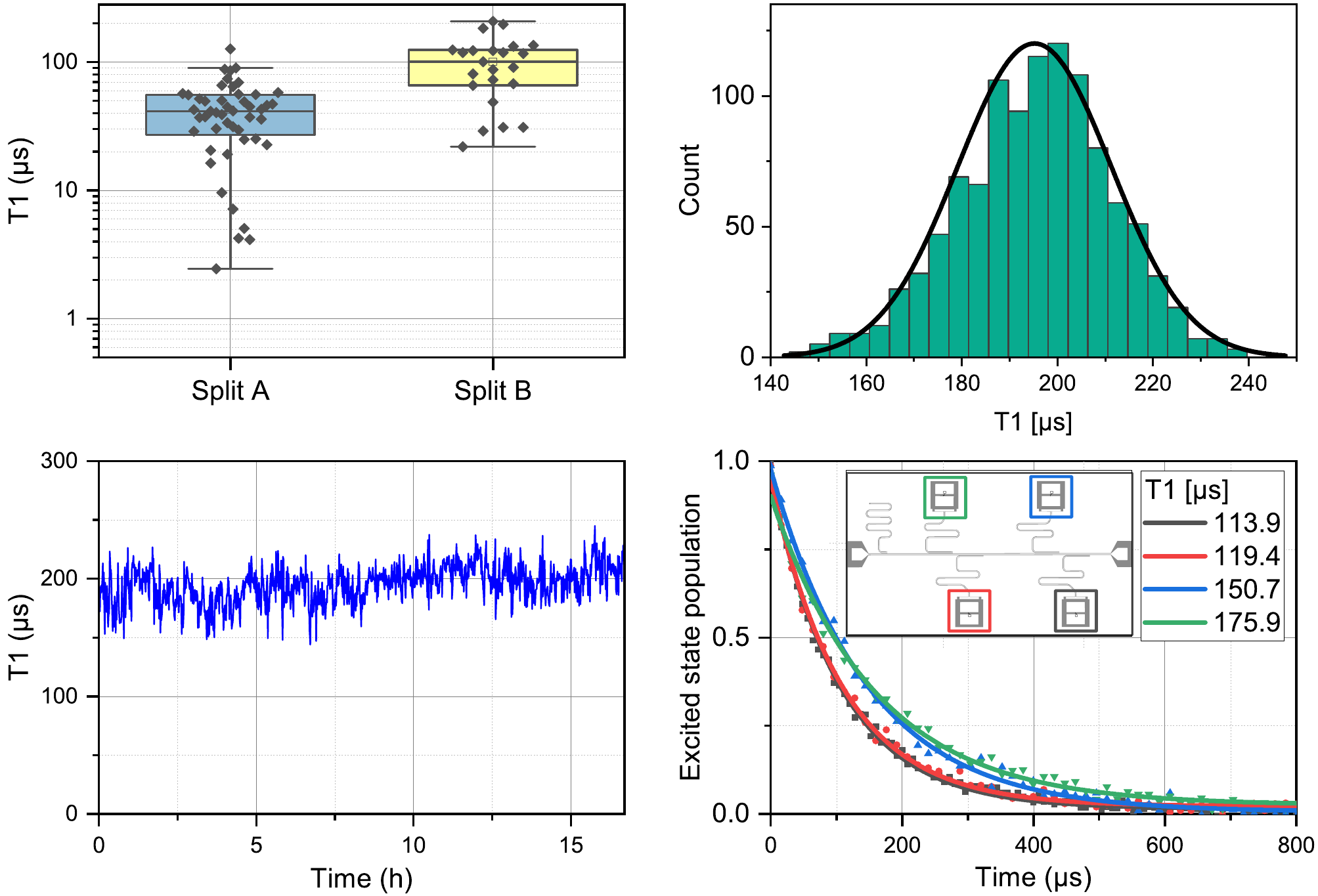}
        \put(9,65){\textbf{a)}}
        \put(60,65){\textbf{b)}}
        \put(9,30.5){\textbf{c)}}
        \put(60,30.5){\textbf{d)}}
    \end{overpic}
    \caption{a) Boxplots of $T_1$ values comparing two split groups in processing. Each data point represents a $T_1$ measurement on a seperate qubit and the vertical line in each box represents the median value of the set. b) $T_1$ histogram of $\approx$1000 measurements on the same qubit. c) Data from plot b) plotted vs. the measurement time to emphasize $T_1$ stability. d) $T_1$ measurements of four qubits on the same chip each exceeding 100\,µs.}
    \label{fig:platzhalter2}
\end{figure*}
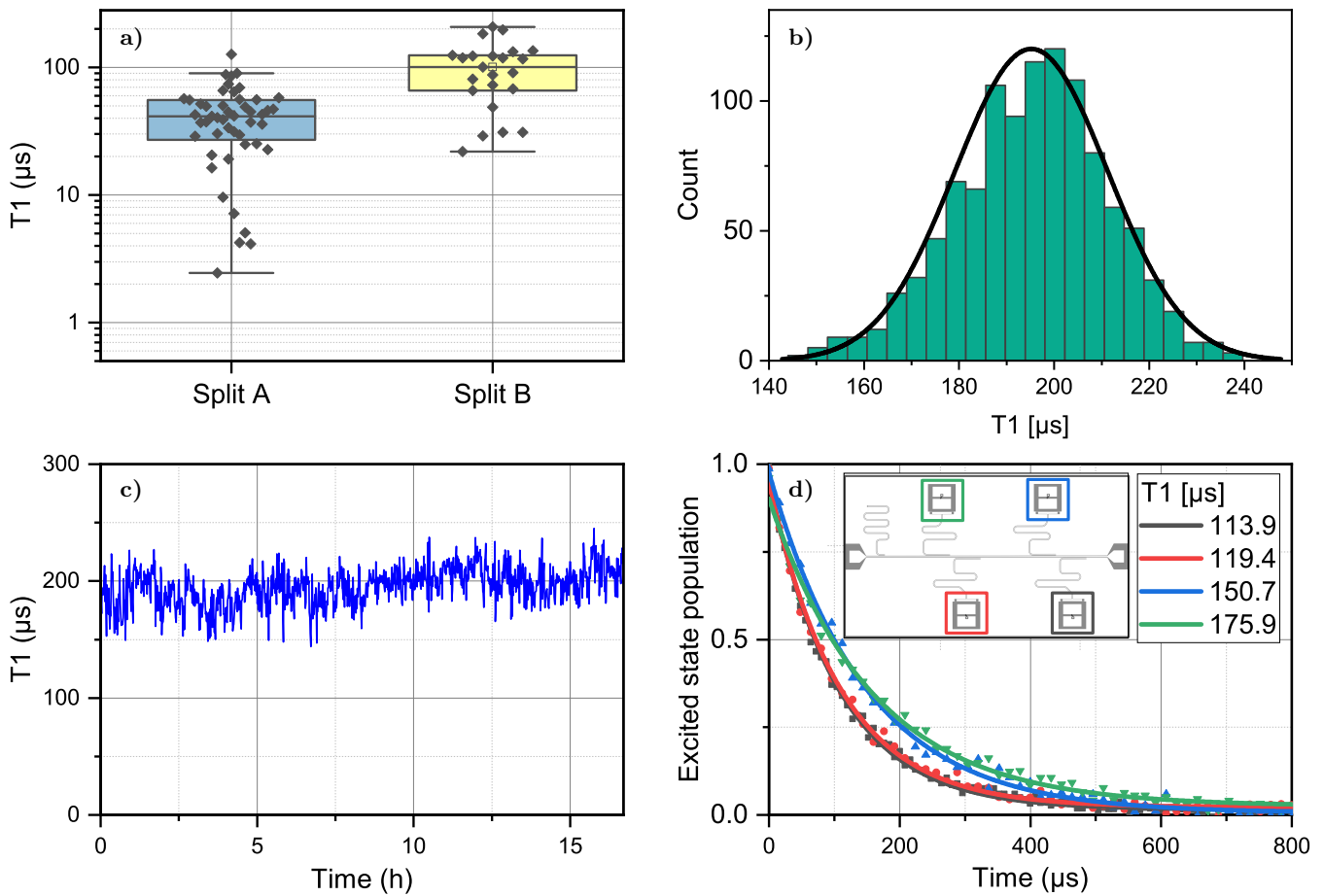

In terms of qubit performance, we will focus in the following on the energy relaxation time $T_1$, as this parameter is typically used as a benchmark for process quality. The presented processing run followed in part our standard fabrication flow, but also involved split groups aiming for additional optimization. Figure \ref{fig:platzhalter2}a) compares two split groups 'A' and 'B' with $T_1$ on a logarithmic scale. Note that in the plot, each data point represents the $T_1$ measurement on a separate qubit and all data generated up to the drafting of the article is shown, nothing is omitted. The qubits were picked from several wafers of the processing run and across the full wafer diameter. While the qubits of split 'A' show a good performance with a median $T_1$ of $\approx \SI{40}{\micro\second}$ (horizontal line in the boxplot) and a single qubit exceeding 100\,µs, there are still several negative outliers with $T_1 < \SI{10}{\micro\second}$. A significant improvement was observed in split group 'B'. Several qubits reached a $T_1$ up to  $\SI{200}{\micro\second}$, accompanied by $T_2^{\mathrm{echo}}$ times above $\SI{200}{\micro\second}$. The $T_1$ median of split 'B' reached a value of $\approx \SI{100}{\micro\second}$. Importantly, while the statistics needs to be increased for a more definite evaluation, the worst qubit of the investigated set still exceeds $\SI{20}{\micro\second}$. Figure \ref{fig:platzhalter2}b) shows a $T_1$ histogram of one of the \textit{'hero'} devices from split 'B' comprised from $\approx$1000 measurement cycles with the mean of the Gaussian fit lying at (196$\pm$16)\,µs. Highlighting the high over-time stability of the qubit, the same data is shown in Fig. \ref{fig:platzhalter2}c), where the 16-hour measurement interval is displayed on the x-axis. During this interval, the energy relaxation time never falls below $\approx \SI{150}{\micro\second}$ and no strong sudden decreases in $T_1$ are observed, which could happen when a strongly coupled two-level system (TLS) with a frequency close to that of the qubit drifts into its vicinity, introducing a substantial energy loss channel. The absence of such drops in $T_1$ is crucial when operating a QPU device to avoid sudden and unforeseeable dips in system performance\footnote{While the data shown in Fig. \ref{fig:platzhalter2}c) displays the result of the whole time interval measured and no cherry-picking of a most favorable interval was applied, the authors acknowledge that for even longer measurement times a strongly-coupled TLS could appear impacting $T_1$ and hence a fair and absolute comparison between qubits, processing approaches, etc. is challenging.}. Smilar to the single-qubit stability, the on-chip quality uniformity has as well to be considered, since the worst performing qubits of a device largely influence the overall system capabilities\cite{mohseni2025buildquantumsupercomputerscaling}. Figure \ref{fig:platzhalter2}d) shows $T_1$ measurements of a chip from split 'B' shown in Fig. \ref{fig:platzhalter2}a), where all qubits on the test chip exceed the threshold of 100\,µs. While the utilized chip design (see inset) is small and only hosts four qubits, this shows that the very high local uniformity of the industry-grade fabrication is also reflected in the qubits' performance.\\
The presented data with limited negative $T_1$ outliers within the set of measured qubits, long-term single-qubit stability, and high on-chip performance homogeneity, shows a promising onset of the possibility to reliably fabricate well-performing, low loss density qubits by the subtractive, industry-grade processing approach.

\section{Summary and Conclusion}
In this article, we presented the results from a CMOS-compatible, industry-grade processing of superconducting qubits on 200\,mm wafers.
By fully automated waferprober characterization of $>10000$ Josephson junctions at room-temperature, we observed a yield of $\SI{99.7}{\percent}$. These measurements can be used as an indicator for qubit yield, as well as a precise tool for qubit frequency prediction. This highly reliable fabrication on large wafers combined with the room-temperature pre-selection capability of devices make our processing a promising approach for scaling the technology to much larger chip sizes with many qubits, or for modular approaches using 3D integration techniques\cite{PhysRevApplied.21.054063,Gold2021-zs,10.1109/MICRO56248.2022.00078,Niu2023-yp}.
Regarding qubit performance, we presented statistical data on energy relaxation times $T_1$ for two split groups within the processing run across many chips and multiple wafers, highlighting the importance of fabrication repeatability and the consideration of negative outliers in the discussion. In split group 'B' we achieved a median $T_1 \approx \SI{100}{\micro\second}$ with the worst qubit lying at $T_1 \approx \SI{20}{\micro\second}$. Some devices in this group reach $T_1 \approx \SI{200}{\micro\second}$ on average, where a long-term measurement over 16 hours revealed consistently good performance. Besides the long-term stability of single qubits, we also stressed the importance of local, on-chip uniformity of qubit quality and presented measurements on a device, where all four qubits exceeded a energy relaxation time of $\SI{100}{\micro\second}$.\\
In conclusion, the presented results demonstrate that an industry-grade, large-wafer processing approach can be reliably and repeatedly used to fabricate high-quality superconducting qubit devices. They also highlight the opportunities of adopting CMOS-type fabrication techniques to address challenges in scaling these devices to the levels required for practical applications.

\section*{Acknowledgments}
The authors would like to thank the Fraunhofer EMFT process engineers and clean room staff for valuable discussion on process development and the professional wafer fabrication.
We acknowledge helpful discussions and assistance with test chip design and setting up of cryogenic measurements from G. Huber, I. Tsitsilin, F. Haslbeck and C. Schneider from the Quantum Computing group at the Walther Meissner Institute.
We would like to thank Z. Tianmu and L. Sigl from Zurich Instruments for their support in setting up the Qubit characterisation software.
This work was funded by the Munich Quantum Valley (MQV) – Consortium Scalable Hardware and Systems Engineering (SHARE), funded by the Bavarian State Government with funds from the Hightech
Agenda Bavaria, the Munich Quantum Valley Quantum Computer Demonstrator - Superconducting Qubits (MUNIQC-SC) 13N16188, funded by the Federal Ministry of Education and Research, Germany, and , and the Open Superconducting Quantum Computers (OpenSuperQPlus) Project - European Quantum Technology Flagship.

\bibliography{paper.bib}

\begin{thebibliography}{17}%
\makeatletter
\providecommand \@ifxundefined [1]{%
 \@ifx{#1\undefined}
}%
\providecommand \@ifnum [1]{%
 \ifnum #1\expandafter \@firstoftwo
 \else \expandafter \@secondoftwo
 \fi
}%
\providecommand \@ifx [1]{%
 \ifx #1\expandafter \@firstoftwo
 \else \expandafter \@secondoftwo
 \fi
}%
\providecommand \natexlab [1]{#1}%
\providecommand \enquote  [1]{``#1''}%
\providecommand \bibnamefont  [1]{#1}%
\providecommand \bibfnamefont [1]{#1}%
\providecommand \citenamefont [1]{#1}%
\providecommand \href@noop [0]{\@secondoftwo}%
\providecommand \href [0]{\begingroup \@sanitize@url \@href}%
\providecommand \@href[1]{\@@startlink{#1}\@@href}%
\providecommand \@@href[1]{\endgroup#1\@@endlink}%
\providecommand \@sanitize@url [0]{\catcode `\\12\catcode `\$12\catcode
  `\&12\catcode `\#12\catcode `\^12\catcode `\_12\catcode `\%12\relax}%
\providecommand \@@startlink[1]{}%
\providecommand \@@endlink[0]{}%
\providecommand \url  [0]{\begingroup\@sanitize@url \@url }%
\providecommand \@url [1]{\endgroup\@href {#1}{\urlprefix }}%
\providecommand \urlprefix  [0]{URL }%
\providecommand \Eprint [0]{\href }%
\providecommand \doibase [0]{https://doi.org/}%
\providecommand \selectlanguage [0]{\@gobble}%
\providecommand \bibinfo  [0]{\@secondoftwo}%
\providecommand \bibfield  [0]{\@secondoftwo}%
\providecommand \translation [1]{[#1]}%
\providecommand \BibitemOpen [0]{}%
\providecommand \bibitemStop [0]{}%
\providecommand \bibitemNoStop [0]{.\EOS\space}%
\providecommand \EOS [0]{\spacefactor3000\relax}%
\providecommand \BibitemShut  [1]{\csname bibitem#1\endcsname}%
\let\auto@bib@innerbib\@empty
\bibitem [{\citenamefont {Mohseni}\ \emph {et~al.}(2025)\citenamefont
  {Mohseni}, \citenamefont {Scherer}, \citenamefont {Johnson}, \citenamefont
  {Wertheim}, \citenamefont {Otten}, \citenamefont {Aadit}, \citenamefont
  {Alexeev}, \citenamefont {Bresniker}, \citenamefont {Camsari}, \citenamefont
  {Chapman}, \citenamefont {Chatterjee}, \citenamefont {Dagnew}, \citenamefont
  {Esposito}, \citenamefont {Fahim}, \citenamefont {Fiorentino}, \citenamefont
  {Gajjar}, \citenamefont {Khalid}, \citenamefont {Kong}, \citenamefont
  {Kulchytskyy}, \citenamefont {Kyoseva}, \citenamefont {Li}, \citenamefont
  {Lott}, \citenamefont {Markov}, \citenamefont {McDermott}, \citenamefont
  {Pedretti}, \citenamefont {Rao}, \citenamefont {Rieffel}, \citenamefont
  {Silva}, \citenamefont {Sorebo}, \citenamefont {Spentzouris}, \citenamefont
  {Steiner}, \citenamefont {Torosov}, \citenamefont {Venturelli}, \citenamefont
  {Visser}, \citenamefont {Webb}, \citenamefont {Zhan}, \citenamefont {Cohen},
  \citenamefont {Ronagh}, \citenamefont {Ho}, \citenamefont {Beausoleil},\ and\
  \citenamefont {Martinis}}]{mohseni2025buildquantumsupercomputerscaling}%
  \BibitemOpen
  \bibfield  {author} {\bibinfo {author} {\bibfnamefont {M.}~\bibnamefont
  {Mohseni}}, \bibinfo {author} {\bibfnamefont {A.}~\bibnamefont {Scherer}},
  \bibinfo {author} {\bibfnamefont {K.~G.}\ \bibnamefont {Johnson}}, \bibinfo
  {author} {\bibfnamefont {O.}~\bibnamefont {Wertheim}}, \bibinfo {author}
  {\bibfnamefont {M.}~\bibnamefont {Otten}}, \bibinfo {author} {\bibfnamefont
  {N.~A.}\ \bibnamefont {Aadit}}, \bibinfo {author} {\bibfnamefont
  {Y.}~\bibnamefont {Alexeev}}, \bibinfo {author} {\bibfnamefont {K.~M.}\
  \bibnamefont {Bresniker}}, \bibinfo {author} {\bibfnamefont {K.~Y.}\
  \bibnamefont {Camsari}}, \bibinfo {author} {\bibfnamefont {B.}~\bibnamefont
  {Chapman}}, \bibinfo {author} {\bibfnamefont {S.}~\bibnamefont {Chatterjee}},
  \bibinfo {author} {\bibfnamefont {G.~A.}\ \bibnamefont {Dagnew}}, \bibinfo
  {author} {\bibfnamefont {A.}~\bibnamefont {Esposito}}, \bibinfo {author}
  {\bibfnamefont {F.}~\bibnamefont {Fahim}}, \bibinfo {author} {\bibfnamefont
  {M.}~\bibnamefont {Fiorentino}}, \bibinfo {author} {\bibfnamefont
  {A.}~\bibnamefont {Gajjar}}, \bibinfo {author} {\bibfnamefont
  {A.}~\bibnamefont {Khalid}}, \bibinfo {author} {\bibfnamefont
  {X.}~\bibnamefont {Kong}}, \bibinfo {author} {\bibfnamefont {B.}~\bibnamefont
  {Kulchytskyy}}, \bibinfo {author} {\bibfnamefont {E.}~\bibnamefont
  {Kyoseva}}, \bibinfo {author} {\bibfnamefont {R.}~\bibnamefont {Li}},
  \bibinfo {author} {\bibfnamefont {P.~A.}\ \bibnamefont {Lott}}, \bibinfo
  {author} {\bibfnamefont {I.~L.}\ \bibnamefont {Markov}}, \bibinfo {author}
  {\bibfnamefont {R.~F.}\ \bibnamefont {McDermott}}, \bibinfo {author}
  {\bibfnamefont {G.}~\bibnamefont {Pedretti}}, \bibinfo {author}
  {\bibfnamefont {P.}~\bibnamefont {Rao}}, \bibinfo {author} {\bibfnamefont
  {E.}~\bibnamefont {Rieffel}}, \bibinfo {author} {\bibfnamefont
  {A.}~\bibnamefont {Silva}}, \bibinfo {author} {\bibfnamefont
  {J.}~\bibnamefont {Sorebo}}, \bibinfo {author} {\bibfnamefont
  {P.}~\bibnamefont {Spentzouris}}, \bibinfo {author} {\bibfnamefont
  {Z.}~\bibnamefont {Steiner}}, \bibinfo {author} {\bibfnamefont
  {B.}~\bibnamefont {Torosov}}, \bibinfo {author} {\bibfnamefont
  {D.}~\bibnamefont {Venturelli}}, \bibinfo {author} {\bibfnamefont {R.~J.}\
  \bibnamefont {Visser}}, \bibinfo {author} {\bibfnamefont {Z.}~\bibnamefont
  {Webb}}, \bibinfo {author} {\bibfnamefont {X.}~\bibnamefont {Zhan}}, \bibinfo
  {author} {\bibfnamefont {Y.}~\bibnamefont {Cohen}}, \bibinfo {author}
  {\bibfnamefont {P.}~\bibnamefont {Ronagh}}, \bibinfo {author} {\bibfnamefont
  {A.}~\bibnamefont {Ho}}, \bibinfo {author} {\bibfnamefont {R.~G.}\
  \bibnamefont {Beausoleil}},\ and\ \bibinfo {author} {\bibfnamefont {J.~M.}\
  \bibnamefont {Martinis}},\ }\href {https://arxiv.org/abs/2411.10406}
  {\bibinfo {title} {How to build a quantum supercomputer: Scaling from
  hundreds to millions of qubits}} (\bibinfo {year} {2025}),\ \Eprint
  {https://arxiv.org/abs/2411.10406} {arXiv:2411.10406} \BibitemShut {NoStop}%
\bibitem [{\citenamefont {Ke}\ \emph {et~al.}(2025)\citenamefont {Ke},
  \citenamefont {Tsai}, \citenamefont {Chen}, \citenamefont {Xu}, \citenamefont
  {Blackwell}, \citenamefont {Snyder}, \citenamefont {Weeden}, \citenamefont
  {Chen}, \citenamefont {Lai}, \citenamefont {Sheu}, \citenamefont {Yang},
  \citenamefont {Wu}, \citenamefont {Ho}, \citenamefont {McDermott},
  \citenamefont {Martinis},\ and\ \citenamefont
  {Chen}}]{ke2025scaffoldassistedwindowjunctionssuperconducting}%
  \BibitemOpen
  \bibfield  {author} {\bibinfo {author} {\bibfnamefont {C.-T.}\ \bibnamefont
  {Ke}}, \bibinfo {author} {\bibfnamefont {J.-Y.}\ \bibnamefont {Tsai}},
  \bibinfo {author} {\bibfnamefont {Y.-C.}\ \bibnamefont {Chen}}, \bibinfo
  {author} {\bibfnamefont {Z.-W.}\ \bibnamefont {Xu}}, \bibinfo {author}
  {\bibfnamefont {E.}~\bibnamefont {Blackwell}}, \bibinfo {author}
  {\bibfnamefont {M.~A.}\ \bibnamefont {Snyder}}, \bibinfo {author}
  {\bibfnamefont {S.}~\bibnamefont {Weeden}}, \bibinfo {author} {\bibfnamefont
  {P.-S.}\ \bibnamefont {Chen}}, \bibinfo {author} {\bibfnamefont {C.-M.}\
  \bibnamefont {Lai}}, \bibinfo {author} {\bibfnamefont {S.-S.}\ \bibnamefont
  {Sheu}}, \bibinfo {author} {\bibfnamefont {Z.}~\bibnamefont {Yang}}, \bibinfo
  {author} {\bibfnamefont {C.-S.}\ \bibnamefont {Wu}}, \bibinfo {author}
  {\bibfnamefont {A.}~\bibnamefont {Ho}}, \bibinfo {author} {\bibfnamefont
  {R.}~\bibnamefont {McDermott}}, \bibinfo {author} {\bibfnamefont
  {J.}~\bibnamefont {Martinis}},\ and\ \bibinfo {author} {\bibfnamefont
  {C.-D.}\ \bibnamefont {Chen}},\ }\href {https://arxiv.org/abs/2503.11010}
  {\bibinfo {title} {Scaffold-assisted window junctions for superconducting
  qubit fabrication}} (\bibinfo {year} {2025}),\ \Eprint
  {https://arxiv.org/abs/2503.11010} {arXiv:2503.11010} \BibitemShut {NoStop}%
\bibitem [{\citenamefont {Megrant}\ and\ \citenamefont
  {Chen}(2025)}]{Megrant2025-si}%
  \BibitemOpen
  \bibfield  {author} {\bibinfo {author} {\bibfnamefont {A.}~\bibnamefont
  {Megrant}}\ and\ \bibinfo {author} {\bibfnamefont {Y.}~\bibnamefont {Chen}},\
  }\bibfield  {title} {\bibinfo {title} {Scaling up superconducting quantum
  computers},\ }\href@noop {} {\bibfield  {journal} {\bibinfo  {journal} {Nat.
  Electron.}\ } (\bibinfo {year} {2025})}\BibitemShut {NoStop}%
\bibitem [{\citenamefont {Lang}\ \emph
  {et~al.}(2025{\natexlab{a}})\citenamefont {Lang}, \citenamefont {Mayer},
  \citenamefont {Weber}, \citenamefont {Dhieb}, \citenamefont {Eisele},
  \citenamefont {Lerch}, \citenamefont {Luo}, \citenamefont {Guizan},
  \citenamefont {Music}, \citenamefont {Sturm-Rogon}, \citenamefont {Zahn},
  \citenamefont {Pereira},\ and\ \citenamefont
  {Kutter}}]{lang2025advancingsuperconductingqubitscmoscompatible}%
  \BibitemOpen
  \bibfield  {author} {\bibinfo {author} {\bibfnamefont {S.~J.~K.}\
  \bibnamefont {Lang}}, \bibinfo {author} {\bibfnamefont {T.}~\bibnamefont
  {Mayer}}, \bibinfo {author} {\bibfnamefont {J.}~\bibnamefont {Weber}},
  \bibinfo {author} {\bibfnamefont {C.}~\bibnamefont {Dhieb}}, \bibinfo
  {author} {\bibfnamefont {I.}~\bibnamefont {Eisele}}, \bibinfo {author}
  {\bibfnamefont {W.}~\bibnamefont {Lerch}}, \bibinfo {author} {\bibfnamefont
  {Z.}~\bibnamefont {Luo}}, \bibinfo {author} {\bibfnamefont {C.~M.}\
  \bibnamefont {Guizan}}, \bibinfo {author} {\bibfnamefont {E.}~\bibnamefont
  {Music}}, \bibinfo {author} {\bibfnamefont {L.}~\bibnamefont {Sturm-Rogon}},
  \bibinfo {author} {\bibfnamefont {D.}~\bibnamefont {Zahn}}, \bibinfo {author}
  {\bibfnamefont {R.~N.}\ \bibnamefont {Pereira}},\ and\ \bibinfo {author}
  {\bibfnamefont {C.}~\bibnamefont {Kutter}},\ }\href
  {https://arxiv.org/abs/2504.18173} {\bibinfo {title} {Advancing
  superconducting qubits: Cmos-compatible processing and room temperature
  characterization for scalable quantum computing beyond 2d architectures}}
  (\bibinfo {year} {2025}{\natexlab{a}}),\ \Eprint
  {https://arxiv.org/abs/2504.18173} {arXiv:2504.18173} \BibitemShut {NoStop}%
\bibitem [{\citenamefont {Van~Damme}\ \emph {et~al.}(2024)\citenamefont
  {Van~Damme}, \citenamefont {Massar}, \citenamefont {Acharya}, \citenamefont
  {Ivanov}, \citenamefont {Perez~Lozano}, \citenamefont {Canvel}, \citenamefont
  {Demarets}, \citenamefont {Vangoidsenhoven}, \citenamefont {Hermans},
  \citenamefont {Lai}, \citenamefont {Vadiraj}, \citenamefont {Mongillo},
  \citenamefont {Wan}, \citenamefont {De~Boeck}, \citenamefont {Poto{\v
  c}nik},\ and\ \citenamefont {De~Greve}}]{Van_Damme2024-vv}%
  \BibitemOpen
  \bibfield  {author} {\bibinfo {author} {\bibfnamefont {J.}~\bibnamefont
  {Van~Damme}}, \bibinfo {author} {\bibfnamefont {S.}~\bibnamefont {Massar}},
  \bibinfo {author} {\bibfnamefont {R.}~\bibnamefont {Acharya}}, \bibinfo
  {author} {\bibfnamefont {T.}~\bibnamefont {Ivanov}}, \bibinfo {author}
  {\bibfnamefont {D.}~\bibnamefont {Perez~Lozano}}, \bibinfo {author}
  {\bibfnamefont {Y.}~\bibnamefont {Canvel}}, \bibinfo {author} {\bibfnamefont
  {M.}~\bibnamefont {Demarets}}, \bibinfo {author} {\bibfnamefont
  {D.}~\bibnamefont {Vangoidsenhoven}}, \bibinfo {author} {\bibfnamefont
  {Y.}~\bibnamefont {Hermans}}, \bibinfo {author} {\bibfnamefont {J.~G.}\
  \bibnamefont {Lai}}, \bibinfo {author} {\bibfnamefont {A.~M.}\ \bibnamefont
  {Vadiraj}}, \bibinfo {author} {\bibfnamefont {M.}~\bibnamefont {Mongillo}},
  \bibinfo {author} {\bibfnamefont {D.}~\bibnamefont {Wan}}, \bibinfo {author}
  {\bibfnamefont {J.}~\bibnamefont {De~Boeck}}, \bibinfo {author}
  {\bibfnamefont {A.}~\bibnamefont {Poto{\v c}nik}},\ and\ \bibinfo {author}
  {\bibfnamefont {K.}~\bibnamefont {De~Greve}},\ }\bibfield  {title} {\bibinfo
  {title} {Advanced {CMOS} manufacturing of superconducting qubits on 300 mm
  wafers},\ }\href@noop {} {\bibfield  {journal} {\bibinfo  {journal} {Nature}\
  }\textbf {\bibinfo {volume} {634}},\ \bibinfo {pages} {74} (\bibinfo {year}
  {2024})}\BibitemShut {NoStop}%
\bibitem [{\citenamefont {Lang}\ \emph {et~al.}(2023)\citenamefont {Lang},
  \citenamefont {Schewski}, \citenamefont {Eisele}, \citenamefont {Kutter},\
  and\ \citenamefont {Lerch}}]{Lang2023-hi}%
  \BibitemOpen
  \bibfield  {author} {\bibinfo {author} {\bibfnamefont {S.}~\bibnamefont
  {Lang}}, \bibinfo {author} {\bibfnamefont {A.}~\bibnamefont {Schewski}},
  \bibinfo {author} {\bibfnamefont {I.}~\bibnamefont {Eisele}}, \bibinfo
  {author} {\bibfnamefont {C.}~\bibnamefont {Kutter}},\ and\ \bibinfo {author}
  {\bibfnamefont {W.}~\bibnamefont {Lerch}},\ }\bibfield  {title} {\bibinfo
  {title} {Aluminum josephson junction formation on 200mm wafers using
  different oxidation techniques},\ }\href@noop {} {\bibfield  {journal}
  {\bibinfo  {journal} {ECS Trans.}\ }\textbf {\bibinfo {volume} {111}},\
  \bibinfo {pages} {41} (\bibinfo {year} {2023})}\BibitemShut {NoStop}%
\bibitem [{\citenamefont {Lang}\ \emph
  {et~al.}(2025{\natexlab{b}})\citenamefont {Lang}, \citenamefont {Eisele},
  \citenamefont {Weber}, \citenamefont {Schewski}, \citenamefont {Lerch},
  \citenamefont {Pereira},\ and\ \citenamefont
  {Kutter}}]{lang2025waferscalecharacterizationalalxoyaljosephson}%
  \BibitemOpen
  \bibfield  {author} {\bibinfo {author} {\bibfnamefont {S.~J.~K.}\
  \bibnamefont {Lang}}, \bibinfo {author} {\bibfnamefont {I.}~\bibnamefont
  {Eisele}}, \bibinfo {author} {\bibfnamefont {J.}~\bibnamefont {Weber}},
  \bibinfo {author} {\bibfnamefont {A.}~\bibnamefont {Schewski}}, \bibinfo
  {author} {\bibfnamefont {W.}~\bibnamefont {Lerch}}, \bibinfo {author}
  {\bibfnamefont {R.~N.}\ \bibnamefont {Pereira}},\ and\ \bibinfo {author}
  {\bibfnamefont {C.}~\bibnamefont {Kutter}},\ }\href
  {https://arxiv.org/abs/2504.16686} {\bibinfo {title} {Wafer-scale
  characterization of al/alxoy/al josephson junctions at room temperature}}
  (\bibinfo {year} {2025}{\natexlab{b}}),\ \Eprint
  {https://arxiv.org/abs/2504.16686} {arXiv:2504.16686 [quant-ph]} \BibitemShut
  {NoStop}%
\bibitem [{\citenamefont {Mayer}\ \emph {et~al.}(2025)\citenamefont {Mayer},
  \citenamefont {Bender}, \citenamefont {Lang}, \citenamefont {Luo},
  \citenamefont {Weber}, \citenamefont {Guizan}, \citenamefont {Dhieb},
  \citenamefont {Zahn}, \citenamefont {Schwarzenbach}, \citenamefont {Hell},
  \citenamefont {Andronic}, \citenamefont {Drost}, \citenamefont {Neumeier},
  \citenamefont {Lerch}, \citenamefont {Nebrich}, \citenamefont {Hagelauer},
  \citenamefont {Eisele}, \citenamefont {Pereira},\ and\ \citenamefont
  {Kutter}}]{mayer20253dintegratedsuperconductingqubitscmoscompatible}%
  \BibitemOpen
  \bibfield  {author} {\bibinfo {author} {\bibfnamefont {T.}~\bibnamefont
  {Mayer}}, \bibinfo {author} {\bibfnamefont {H.}~\bibnamefont {Bender}},
  \bibinfo {author} {\bibfnamefont {S.~J.~K.}\ \bibnamefont {Lang}}, \bibinfo
  {author} {\bibfnamefont {Z.}~\bibnamefont {Luo}}, \bibinfo {author}
  {\bibfnamefont {J.}~\bibnamefont {Weber}}, \bibinfo {author} {\bibfnamefont
  {C.~M.}\ \bibnamefont {Guizan}}, \bibinfo {author} {\bibfnamefont
  {C.}~\bibnamefont {Dhieb}}, \bibinfo {author} {\bibfnamefont
  {D.}~\bibnamefont {Zahn}}, \bibinfo {author} {\bibfnamefont {L.}~\bibnamefont
  {Schwarzenbach}}, \bibinfo {author} {\bibfnamefont {W.}~\bibnamefont {Hell}},
  \bibinfo {author} {\bibfnamefont {M.}~\bibnamefont {Andronic}}, \bibinfo
  {author} {\bibfnamefont {A.}~\bibnamefont {Drost}}, \bibinfo {author}
  {\bibfnamefont {K.}~\bibnamefont {Neumeier}}, \bibinfo {author}
  {\bibfnamefont {W.}~\bibnamefont {Lerch}}, \bibinfo {author} {\bibfnamefont
  {L.}~\bibnamefont {Nebrich}}, \bibinfo {author} {\bibfnamefont
  {A.}~\bibnamefont {Hagelauer}}, \bibinfo {author} {\bibfnamefont
  {I.}~\bibnamefont {Eisele}}, \bibinfo {author} {\bibfnamefont {R.~N.}\
  \bibnamefont {Pereira}},\ and\ \bibinfo {author} {\bibfnamefont
  {C.}~\bibnamefont {Kutter}},\ }\href {https://arxiv.org/abs/2505.04337}
  {\bibinfo {title} {3d-integrated superconducting qubits: Cmos-compatible,
  wafer-scale processing for flip-chip architectures}} (\bibinfo {year}
  {2025}),\ \Eprint {https://arxiv.org/abs/2505.04337} {arXiv:2505.04337}
  \BibitemShut {NoStop}%
\bibitem [{\citenamefont {Ambegaokar}\ and\ \citenamefont
  {Baratoff}(1963)}]{ambegaokar_tunneling_1963}%
  \BibitemOpen
  \bibfield  {author} {\bibinfo {author} {\bibfnamefont {V.}~\bibnamefont
  {Ambegaokar}}\ and\ \bibinfo {author} {\bibfnamefont {A.}~\bibnamefont
  {Baratoff}},\ }\bibfield  {title} {\bibinfo {title} {Tunneling {Between}
  {Superconductors}},\ }\href {https://doi.org/10.1103/PhysRevLett.10.486}
  {\bibfield  {journal} {\bibinfo  {journal} {Physical Review Letters}\
  }\textbf {\bibinfo {volume} {10}},\ \bibinfo {pages} {486} (\bibinfo {year}
  {1963})}\BibitemShut {NoStop}%
\bibitem [{\citenamefont {Hertzberg}\ \emph {et~al.}(2021)\citenamefont
  {Hertzberg}, \citenamefont {Zhang}, \citenamefont {Rosenblatt}, \citenamefont
  {Magesan}, \citenamefont {Smolin}, \citenamefont {Yau}, \citenamefont
  {Adiga}, \citenamefont {Sandberg}, \citenamefont {Brink}, \citenamefont
  {Chow},\ and\ \citenamefont {Orcutt}}]{Hertzberg2021-yn}%
  \BibitemOpen
  \bibfield  {author} {\bibinfo {author} {\bibfnamefont {J.~B.}\ \bibnamefont
  {Hertzberg}}, \bibinfo {author} {\bibfnamefont {E.~J.}\ \bibnamefont
  {Zhang}}, \bibinfo {author} {\bibfnamefont {S.}~\bibnamefont {Rosenblatt}},
  \bibinfo {author} {\bibfnamefont {E.}~\bibnamefont {Magesan}}, \bibinfo
  {author} {\bibfnamefont {J.~A.}\ \bibnamefont {Smolin}}, \bibinfo {author}
  {\bibfnamefont {J.-B.}\ \bibnamefont {Yau}}, \bibinfo {author} {\bibfnamefont
  {V.~P.}\ \bibnamefont {Adiga}}, \bibinfo {author} {\bibfnamefont
  {M.}~\bibnamefont {Sandberg}}, \bibinfo {author} {\bibfnamefont
  {M.}~\bibnamefont {Brink}}, \bibinfo {author} {\bibfnamefont {J.~M.}\
  \bibnamefont {Chow}},\ and\ \bibinfo {author} {\bibfnamefont {J.~S.}\
  \bibnamefont {Orcutt}},\ }\bibfield  {title} {\bibinfo {title}
  {Laser-annealing josephson junctions for yielding scaled-up superconducting
  quantum processors},\ }\href@noop {} {\bibfield  {journal} {\bibinfo
  {journal} {Npj Quantum Inf.}\ }\textbf {\bibinfo {volume} {7}} (\bibinfo
  {year} {2021})}\BibitemShut {NoStop}%
\bibitem [{\citenamefont {Zhang}\ \emph {et~al.}(2022)\citenamefont {Zhang},
  \citenamefont {Srinivasan}, \citenamefont {Sundaresan}, \citenamefont
  {Bogorin}, \citenamefont {Martin}, \citenamefont {Hertzberg}, \citenamefont
  {Timmerwilke}, \citenamefont {Pritchett}, \citenamefont {Yau}, \citenamefont
  {Wang}, \citenamefont {Landers}, \citenamefont {Lewandowski}, \citenamefont
  {Narasgond}, \citenamefont {Rosenblatt}, \citenamefont {Keefe}, \citenamefont
  {Lauer}, \citenamefont {Rothwell}, \citenamefont {McClure}, \citenamefont
  {Dial}, \citenamefont {Orcutt}, \citenamefont {Brink},\ and\ \citenamefont
  {Chow}}]{doi:10.1126/sciadv.abi6690}%
  \BibitemOpen
  \bibfield  {author} {\bibinfo {author} {\bibfnamefont {E.~J.}\ \bibnamefont
  {Zhang}}, \bibinfo {author} {\bibfnamefont {S.}~\bibnamefont {Srinivasan}},
  \bibinfo {author} {\bibfnamefont {N.}~\bibnamefont {Sundaresan}}, \bibinfo
  {author} {\bibfnamefont {D.~F.}\ \bibnamefont {Bogorin}}, \bibinfo {author}
  {\bibfnamefont {Y.}~\bibnamefont {Martin}}, \bibinfo {author} {\bibfnamefont
  {J.~B.}\ \bibnamefont {Hertzberg}}, \bibinfo {author} {\bibfnamefont
  {J.}~\bibnamefont {Timmerwilke}}, \bibinfo {author} {\bibfnamefont {E.~J.}\
  \bibnamefont {Pritchett}}, \bibinfo {author} {\bibfnamefont {J.-B.}\
  \bibnamefont {Yau}}, \bibinfo {author} {\bibfnamefont {C.}~\bibnamefont
  {Wang}}, \bibinfo {author} {\bibfnamefont {W.}~\bibnamefont {Landers}},
  \bibinfo {author} {\bibfnamefont {E.~P.}\ \bibnamefont {Lewandowski}},
  \bibinfo {author} {\bibfnamefont {A.}~\bibnamefont {Narasgond}}, \bibinfo
  {author} {\bibfnamefont {S.}~\bibnamefont {Rosenblatt}}, \bibinfo {author}
  {\bibfnamefont {G.~A.}\ \bibnamefont {Keefe}}, \bibinfo {author}
  {\bibfnamefont {I.}~\bibnamefont {Lauer}}, \bibinfo {author} {\bibfnamefont
  {M.~B.}\ \bibnamefont {Rothwell}}, \bibinfo {author} {\bibfnamefont {D.~T.}\
  \bibnamefont {McClure}}, \bibinfo {author} {\bibfnamefont {O.~E.}\
  \bibnamefont {Dial}}, \bibinfo {author} {\bibfnamefont {J.~S.}\ \bibnamefont
  {Orcutt}}, \bibinfo {author} {\bibfnamefont {M.}~\bibnamefont {Brink}},\ and\
  \bibinfo {author} {\bibfnamefont {J.~M.}\ \bibnamefont {Chow}},\ }\bibfield
  {title} {\bibinfo {title} {High-performance superconducting quantum
  processors via laser annealing of transmon qubits},\ }\href
  {https://doi.org/10.1126/sciadv.abi6690} {\bibfield  {journal} {\bibinfo
  {journal} {Science Advances}\ }\textbf {\bibinfo {volume} {8}},\ \bibinfo
  {pages} {eabi6690} (\bibinfo {year} {2022})},\ \Eprint
  {https://arxiv.org/abs/https://www.science.org/doi/pdf/10.1126/sciadv.abi6690}
  {https://www.science.org/doi/pdf/10.1126/sciadv.abi6690} \BibitemShut
  {NoStop}%
\bibitem [{\citenamefont {Kim}\ \emph {et~al.}(2022)\citenamefont {Kim},
  \citenamefont {Jünger}, \citenamefont {Morvan}, \citenamefont {Barnard},
  \citenamefont {Livingston}, \citenamefont {Altoé}, \citenamefont {Kim},
  \citenamefont {Song}, \citenamefont {Chen}, \citenamefont {Kreikebaum},
  \citenamefont {Ogletree}, \citenamefont {Santiago},\ and\ \citenamefont
  {Siddiqi}}]{10.1063/5.0102092}%
  \BibitemOpen
  \bibfield  {author} {\bibinfo {author} {\bibfnamefont {H.}~\bibnamefont
  {Kim}}, \bibinfo {author} {\bibfnamefont {C.}~\bibnamefont {Jünger}},
  \bibinfo {author} {\bibfnamefont {A.}~\bibnamefont {Morvan}}, \bibinfo
  {author} {\bibfnamefont {E.~S.}\ \bibnamefont {Barnard}}, \bibinfo {author}
  {\bibfnamefont {W.~P.}\ \bibnamefont {Livingston}}, \bibinfo {author}
  {\bibfnamefont {M.~V.~P.}\ \bibnamefont {Altoé}}, \bibinfo {author}
  {\bibfnamefont {Y.}~\bibnamefont {Kim}}, \bibinfo {author} {\bibfnamefont
  {C.}~\bibnamefont {Song}}, \bibinfo {author} {\bibfnamefont {L.}~\bibnamefont
  {Chen}}, \bibinfo {author} {\bibfnamefont {J.~M.}\ \bibnamefont
  {Kreikebaum}}, \bibinfo {author} {\bibfnamefont {D.~F.}\ \bibnamefont
  {Ogletree}}, \bibinfo {author} {\bibfnamefont {D.~I.}\ \bibnamefont
  {Santiago}},\ and\ \bibinfo {author} {\bibfnamefont {I.}~\bibnamefont
  {Siddiqi}},\ }\bibfield  {title} {\bibinfo {title} {Effects of
  laser-annealing on fixed-frequency superconducting qubits},\ }\href
  {https://doi.org/10.1063/5.0102092} {\bibfield  {journal} {\bibinfo
  {journal} {Applied Physics Letters}\ }\textbf {\bibinfo {volume} {121}},\
  \bibinfo {pages} {142601} (\bibinfo {year} {2022})},\ \Eprint
  {https://arxiv.org/abs/https://pubs.aip.org/aip/apl/article-pdf/doi/10.1063/5.0102092/16484380/142601\_1\_online.pdf}
  {https://pubs.aip.org/aip/apl/article-pdf/doi/10.1063/5.0102092/16484380/142601\_1\_online.pdf}
  \BibitemShut {NoStop}%
\bibitem [{\citenamefont {Osman}\ \emph {et~al.}(2023)\citenamefont {Osman},
  \citenamefont {Fern\'andez-Pend\'as}, \citenamefont {Warren}, \citenamefont
  {Kosen}, \citenamefont {Scigliuzzo}, \citenamefont {Frisk~Kockum},
  \citenamefont {Tancredi}, \citenamefont {Fadavi~Roudsari},\ and\
  \citenamefont {Bylander}}]{PhysRevResearch.5.043001}%
  \BibitemOpen
  \bibfield  {author} {\bibinfo {author} {\bibfnamefont {A.}~\bibnamefont
  {Osman}}, \bibinfo {author} {\bibfnamefont {J.}~\bibnamefont
  {Fern\'andez-Pend\'as}}, \bibinfo {author} {\bibfnamefont {C.}~\bibnamefont
  {Warren}}, \bibinfo {author} {\bibfnamefont {S.}~\bibnamefont {Kosen}},
  \bibinfo {author} {\bibfnamefont {M.}~\bibnamefont {Scigliuzzo}}, \bibinfo
  {author} {\bibfnamefont {A.}~\bibnamefont {Frisk~Kockum}}, \bibinfo {author}
  {\bibfnamefont {G.}~\bibnamefont {Tancredi}}, \bibinfo {author}
  {\bibfnamefont {A.}~\bibnamefont {Fadavi~Roudsari}},\ and\ \bibinfo {author}
  {\bibfnamefont {J.}~\bibnamefont {Bylander}},\ }\bibfield  {title} {\bibinfo
  {title} {Mitigation of frequency collisions in superconducting quantum
  processors},\ }\href {https://doi.org/10.1103/PhysRevResearch.5.043001}
  {\bibfield  {journal} {\bibinfo  {journal} {Phys. Rev. Res.}\ }\textbf
  {\bibinfo {volume} {5}},\ \bibinfo {pages} {043001} (\bibinfo {year}
  {2023})}\BibitemShut {NoStop}%
\bibitem [{\citenamefont {Field}\ \emph {et~al.}(2024)\citenamefont {Field},
  \citenamefont {Chen}, \citenamefont {Scharmann}, \citenamefont {Sete},
  \citenamefont {Oruc}, \citenamefont {Vu}, \citenamefont {Kosenko},
  \citenamefont {Mutus}, \citenamefont {Poletto},\ and\ \citenamefont
  {Bestwick}}]{PhysRevApplied.21.054063}%
  \BibitemOpen
  \bibfield  {author} {\bibinfo {author} {\bibfnamefont {M.}~\bibnamefont
  {Field}}, \bibinfo {author} {\bibfnamefont {A.~Q.}\ \bibnamefont {Chen}},
  \bibinfo {author} {\bibfnamefont {B.}~\bibnamefont {Scharmann}}, \bibinfo
  {author} {\bibfnamefont {E.~A.}\ \bibnamefont {Sete}}, \bibinfo {author}
  {\bibfnamefont {F.}~\bibnamefont {Oruc}}, \bibinfo {author} {\bibfnamefont
  {K.}~\bibnamefont {Vu}}, \bibinfo {author} {\bibfnamefont {V.}~\bibnamefont
  {Kosenko}}, \bibinfo {author} {\bibfnamefont {J.~Y.}\ \bibnamefont {Mutus}},
  \bibinfo {author} {\bibfnamefont {S.}~\bibnamefont {Poletto}},\ and\ \bibinfo
  {author} {\bibfnamefont {A.}~\bibnamefont {Bestwick}},\ }\bibfield  {title}
  {\bibinfo {title} {Modular superconducting-qubit architecture with a
  multichip tunable coupler},\ }\href
  {https://doi.org/10.1103/PhysRevApplied.21.054063} {\bibfield  {journal}
  {\bibinfo  {journal} {Phys. Rev. Appl.}\ }\textbf {\bibinfo {volume} {21}},\
  \bibinfo {pages} {054063} (\bibinfo {year} {2024})}\BibitemShut {NoStop}%
\bibitem [{\citenamefont {Gold}\ \emph {et~al.}(2021)\citenamefont {Gold},
  \citenamefont {Paquette}, \citenamefont {Stockklauser}, \citenamefont
  {Reagor}, \citenamefont {Alam}, \citenamefont {Bestwick}, \citenamefont
  {Didier}, \citenamefont {Nersisyan}, \citenamefont {Oruc}, \citenamefont
  {Razavi}, \citenamefont {Scharmann}, \citenamefont {Sete}, \citenamefont
  {Sur}, \citenamefont {Venturelli}, \citenamefont {Winkleblack}, \citenamefont
  {Wudarski}, \citenamefont {Harburn},\ and\ \citenamefont
  {Rigetti}}]{Gold2021-zs}%
  \BibitemOpen
  \bibfield  {author} {\bibinfo {author} {\bibfnamefont {A.}~\bibnamefont
  {Gold}}, \bibinfo {author} {\bibfnamefont {J.~P.}\ \bibnamefont {Paquette}},
  \bibinfo {author} {\bibfnamefont {A.}~\bibnamefont {Stockklauser}}, \bibinfo
  {author} {\bibfnamefont {M.~J.}\ \bibnamefont {Reagor}}, \bibinfo {author}
  {\bibfnamefont {M.~S.}\ \bibnamefont {Alam}}, \bibinfo {author}
  {\bibfnamefont {A.}~\bibnamefont {Bestwick}}, \bibinfo {author}
  {\bibfnamefont {N.}~\bibnamefont {Didier}}, \bibinfo {author} {\bibfnamefont
  {A.}~\bibnamefont {Nersisyan}}, \bibinfo {author} {\bibfnamefont
  {F.}~\bibnamefont {Oruc}}, \bibinfo {author} {\bibfnamefont {A.}~\bibnamefont
  {Razavi}}, \bibinfo {author} {\bibfnamefont {B.}~\bibnamefont {Scharmann}},
  \bibinfo {author} {\bibfnamefont {E.~A.}\ \bibnamefont {Sete}}, \bibinfo
  {author} {\bibfnamefont {B.}~\bibnamefont {Sur}}, \bibinfo {author}
  {\bibfnamefont {D.}~\bibnamefont {Venturelli}}, \bibinfo {author}
  {\bibfnamefont {C.~J.}\ \bibnamefont {Winkleblack}}, \bibinfo {author}
  {\bibfnamefont {F.}~\bibnamefont {Wudarski}}, \bibinfo {author}
  {\bibfnamefont {M.}~\bibnamefont {Harburn}},\ and\ \bibinfo {author}
  {\bibfnamefont {C.}~\bibnamefont {Rigetti}},\ }\bibfield  {title} {\bibinfo
  {title} {Entanglement across separate silicon dies in a modular
  superconducting qubit device},\ }\href@noop {} {\bibfield  {journal}
  {\bibinfo  {journal} {Npj Quantum Inf.}\ }\textbf {\bibinfo {volume} {7}}
  (\bibinfo {year} {2021})}\BibitemShut {NoStop}%
\bibitem [{\citenamefont {Smith}\ \emph {et~al.}(2023)\citenamefont {Smith},
  \citenamefont {Ravi}, \citenamefont {Baker},\ and\ \citenamefont
  {Chong}}]{10.1109/MICRO56248.2022.00078}%
  \BibitemOpen
  \bibfield  {author} {\bibinfo {author} {\bibfnamefont {K.~N.}\ \bibnamefont
  {Smith}}, \bibinfo {author} {\bibfnamefont {G.~S.}\ \bibnamefont {Ravi}},
  \bibinfo {author} {\bibfnamefont {J.~M.}\ \bibnamefont {Baker}},\ and\
  \bibinfo {author} {\bibfnamefont {F.~T.}\ \bibnamefont {Chong}},\ }\bibfield
  {title} {\bibinfo {title} {Scaling superconducting quantum computers with
  chiplet architectures},\ }in\ \href
  {https://doi.org/10.1109/MICRO56248.2022.00078} {\emph {\bibinfo {booktitle}
  {Proceedings of the 55th Annual IEEE/ACM International Symposium on
  Microarchitecture}}},\ \bibinfo {series and number} {MICRO '22}\ (\bibinfo
  {publisher} {IEEE Press},\ \bibinfo {year} {2023})\ p.\ \bibinfo {pages}
  {1092–1109}\BibitemShut {NoStop}%
\bibitem [{\citenamefont {Niu}\ \emph {et~al.}(2023)\citenamefont {Niu},
  \citenamefont {Zhang}, \citenamefont {Liu}, \citenamefont {Qiu},
  \citenamefont {Huang}, \citenamefont {Huang}, \citenamefont {Jia},
  \citenamefont {Liu}, \citenamefont {Tao}, \citenamefont {Wei}, \citenamefont
  {Zhou}, \citenamefont {Zou}, \citenamefont {Chen}, \citenamefont {Deng},
  \citenamefont {Deng}, \citenamefont {Hu}, \citenamefont {Hu}, \citenamefont
  {Li}, \citenamefont {Tan}, \citenamefont {Xu}, \citenamefont {Yan},
  \citenamefont {Yan}, \citenamefont {Liu}, \citenamefont {Zhong},
  \citenamefont {Cleland},\ and\ \citenamefont {Yu}}]{Niu2023-yp}%
  \BibitemOpen
  \bibfield  {author} {\bibinfo {author} {\bibfnamefont {J.}~\bibnamefont
  {Niu}}, \bibinfo {author} {\bibfnamefont {L.}~\bibnamefont {Zhang}}, \bibinfo
  {author} {\bibfnamefont {Y.}~\bibnamefont {Liu}}, \bibinfo {author}
  {\bibfnamefont {J.}~\bibnamefont {Qiu}}, \bibinfo {author} {\bibfnamefont
  {W.}~\bibnamefont {Huang}}, \bibinfo {author} {\bibfnamefont
  {J.}~\bibnamefont {Huang}}, \bibinfo {author} {\bibfnamefont
  {H.}~\bibnamefont {Jia}}, \bibinfo {author} {\bibfnamefont {J.}~\bibnamefont
  {Liu}}, \bibinfo {author} {\bibfnamefont {Z.}~\bibnamefont {Tao}}, \bibinfo
  {author} {\bibfnamefont {W.}~\bibnamefont {Wei}}, \bibinfo {author}
  {\bibfnamefont {Y.}~\bibnamefont {Zhou}}, \bibinfo {author} {\bibfnamefont
  {W.}~\bibnamefont {Zou}}, \bibinfo {author} {\bibfnamefont {Y.}~\bibnamefont
  {Chen}}, \bibinfo {author} {\bibfnamefont {X.}~\bibnamefont {Deng}}, \bibinfo
  {author} {\bibfnamefont {X.}~\bibnamefont {Deng}}, \bibinfo {author}
  {\bibfnamefont {C.}~\bibnamefont {Hu}}, \bibinfo {author} {\bibfnamefont
  {L.}~\bibnamefont {Hu}}, \bibinfo {author} {\bibfnamefont {J.}~\bibnamefont
  {Li}}, \bibinfo {author} {\bibfnamefont {D.}~\bibnamefont {Tan}}, \bibinfo
  {author} {\bibfnamefont {Y.}~\bibnamefont {Xu}}, \bibinfo {author}
  {\bibfnamefont {F.}~\bibnamefont {Yan}}, \bibinfo {author} {\bibfnamefont
  {T.}~\bibnamefont {Yan}}, \bibinfo {author} {\bibfnamefont {S.}~\bibnamefont
  {Liu}}, \bibinfo {author} {\bibfnamefont {Y.}~\bibnamefont {Zhong}}, \bibinfo
  {author} {\bibfnamefont {A.~N.}\ \bibnamefont {Cleland}},\ and\ \bibinfo
  {author} {\bibfnamefont {D.}~\bibnamefont {Yu}},\ }\bibfield  {title}
  {\bibinfo {title} {Low-loss interconnects for modular superconducting quantum
  processors},\ }\href@noop {} {\bibfield  {journal} {\bibinfo  {journal} {Nat.
  Electron.}\ }\textbf {\bibinfo {volume} {6}},\ \bibinfo {pages} {235}
  (\bibinfo {year} {2023})}\BibitemShut {NoStop}%
\end{thebibliography}%


%
\end{document}